\documentclass{sig-alternate-10pt}
\usepackage{comment}
\usepackage{graphicx}
\usepackage{url} 
\begin{document}
 
\title{Prediction of Cyberbullying Incidents on the Instagram Social Network}
 
	\author{Homa Hosseinmardi$^\ddag$, Sabrina Arredondo Mattson$^\S$, Rahat Ibn Rafiq$^\ddag$,\\ 
 		 Richard Han$^\ddag$, Qin Lv$^\ddag$, Shivakant Mishra$^\ddag$ \\ 
 		\affaddr{$^\ddag$Computer Science Department, $^\S$Institute of Behavioral Science} \\ 
 		\affaddr{University of Colorado Boulder, Boulder, CO 80309 USA} \\ 
\email{homa.hosseinmardi@colorado.edu}
}

\maketitle
\begin{abstract}
Cyberbullying is a growing problem affecting more than half of all American teens. The main goal of this paper is to investigate fundamentally new approaches to understand and automatically detect and predict incidents of cyberbullying in Instagram, a media-based mobile social network. In this work,
we have collected a sample data set consisting of Instagram images and their associated comments.  We then designed a labeling study and employed human contributors at the crowd-sourced CrowdFlower website to label these media sessions for cyberbullying.  A detailed analysis of the labeled data is then presented, including a study of relationships between cyberbullying and a host of features such as cyberaggression, profanity, social graph features, temporal commenting behavior, linguistic content, and image content. Using the labeled data, we further design and evaluate the performance of classifiers to automatically detect and predict incidents of cyberbullying and cyberaggression.
\end{abstract}
 




\section{Introduction}
\label{sec:intro} 

As online social networks (OSNs) have grown in popularity, instances of cyberbullying in OSNs have become an increasing concern.  In fact more than half of American teens have been the victims of cyberbullying \cite{stop}. Although cyberbullying may not cause any physical damage initially, it has potentially devastating psychological effects like depression, low self-esteem, suicide ideation, and even suicide \cite{suicideidea,Hinduja,Menesini,van2015automatic} and can have long term effects in the future life of victims \cite{cyberCNN}. Incidents of cyberbullying with extreme consequences such as suicide are now routinely reported in the popular press. For example cyberbullying of Jessica Logan via her image shared in Facebook and MySpace and of Hope Sitwell with her image shared in MySpace is attributed to their suicides \cite{logan}, \cite{imagebully}. Also, nine teenagers reportedly committed suicide having experienced cyberbullying in Ask.fm \cite{askfmsuicides}. Although cyberbullying is not the direct cause of these suicides,  it was viewed as a contributing factor in the death of these teenagers \cite{cyber1}. 

Given the gravity of the consequences cyberbullying has on its victims and its rapid spread among middle and high school students, there is an immediate and pressing need for research to understand how cyberbullying occurs in OSNs today, so that effective techniques can be developed to accurately detect cyberbullying. In \cite{van2015automatic}, it is reported that experts in the field of cyberbullying could favor automatic monitoring of cyberbullying on social networking sites and propose effective follow-up strategies.

Our work makes the important distinction between {\em cyberaggression},
which concerns the aggressiveness of a single remark toward a user,
and {\em cyberbullying}, which concerns the overall pattern of aggressiveness of many remarks directed at a user.  It is this pattern of aggression that severely impacts many teens.  In fact, cyberbullying has been defined as intentionally aggressive behavior that is \emph{repeatedly carried out} in an online context against a person \emph{who cannot easily defend him or herself} \cite{kowalski2012cyberbullying,patchin2012update}.  It is important to this definition of cyberbullying that both the frequency of negativity and the imbalance of power between the victim and perpetrator be taken into account.  In contrast, cyberaggression is defined as an instance of using digital media to intentionally harm another person~\cite{kowalski2012cyberbullying}.

\begin{figure}[!ht]
\centering
\includegraphics[width=0.3\textwidth]{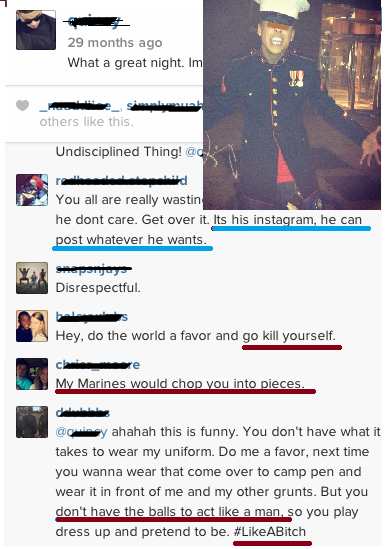}
\caption{An example of comments posted on Instagram.  To give more room for the text, we have moved the associated image to overlay some of the text.}
\label{fig:insta}
\end{figure}

Facebook, Twitter, YouTube, Ask.fm, and Instagram have been listed as the top five networks with the highest percentage of users reporting experience of cyberbullying \cite{annual}.  Instagram is of particular interest as it is a media-based mobile social network, which allows  users to post and comment on images.  Cyberbullying in Instagram can happen in different ways, including posting a humiliating image of someone else by perhaps editing the image, posting mean or hateful comments, aggressive captions or hashtags, or creating fake profiles pretending to be someone else \cite{cyberinsta}. Figure \ref{fig:insta} illustrates an example of an attack in Instagram in which hateful comments were posted for the profile owner.

The main goal of this paper is to first study cyberbullying in Instagram and 
then develop classifiers to automatically detect cyberbullying incidents
and predict onset of cyberbullying incidents. To do so, we first collected a
large sample of Instagram data comprised of 3,165K media sessions (images and their associated comments) taken from 25K user profiles.
Next, we designed and conducted a set of labeling surveys using
human labelers at the crowd-sourced CrowdFlower website to identify occurrences of cyberbullying and cyberaggression in Instagram.  We then analyzed the labeled data set, reporting different features of these media sessions with respect to cyberbullying. Finally, based on our analysis, we have designed and
evaluated classifiers for detecting cyberbullying incidents and predicting onset
of cyberbullying incidents. This paper make the following important 
contributions:

\begin{itemize}
	\item We provide a clear distinction between cyberbullying and general
cyberaggression. Cyberbullying is a stricter form of cyberaggression, 
while most of the earlier research in this area has focused on identifying
cyberaggression.
	\item We obtain ground truth about cyberbullying behavior in Instagram by asking human crowd-sourcers to label Instagram images and their associated comments according to both the more restrictive definition of cyberbullying and the more general definition of cyberaggression. Labelers  are provided with the image and its associated comments at the same time to be able to understand the context and label accordingly. 
	\item We present a novel detailed analysis of the labeled media sessions, including the relationships between cyberbullying and a host of factors, such as cyberaggression, profanity, social graph properties (liking, followers/following), the interarrival time of comments, the linguistic content of comments, and labeled image content. 
	\item We design and evaluate multi-modal classifiers that extend beyond merely the text dimension for detecting cyberbullying and further incorporate image-based features and user statistics in the detection. 
	\item Finally, we design and evaluate multi-modal classifiers to predict cyberbullying based on the labeled data. To the best of our knowledge, no other research has addressed the problem of predicting cyberbullying in social networks. 
\end{itemize}

\section{Related Works}
\label{sec:related}
Prior works that investigated cyberbullying \cite{Ptaszynski2010,Improved,usingML,Dinakar_modelingthe2,Twitter,kontostathis2013detecting,xu2012learning,Nahar14,Nahar13,dinakar2011modeling,nahar2012sentiment} are more accurately described as research on cyberaggression, since these works do not take into account both the frequency of aggression and the imbalance of power.  These works have largely applied a text analysis approach to online comments, since this approach results in higher precision and lower false positives than simpler list-based matching of profane words \cite{sood2012profanity}.  Previous research \cite{nandhini2015cyberbullying,nandhini2015online,kontostathis2013detecting,Nahar14} applied text based cyberbullying on Formspring.me and Myspace dataset. Dinakar {\em et al.} investigated both explicit and implicit cyberbullying by analyzing negative text comments on YouTube and Formspring profiles~\cite{Dinakar_modelingthe2}. Sanchez and Kumar proposed using a Naive Bayes classifier to find inappropriate words in Twitter text data for bullying detection~\cite{Twitter}. They tracked  potential bullies, their followers, and the victims. Also some researchers tried to detect bullies and victims by looking at the number of received and sent, beside detecting aggressive comments \cite{nalini2015classification} and \cite{nahar2012sentiment} . Dadvar {\em et al.} investigated how combining text analysis with MySpace user profile information such as gender can improve the accuracy of cyberbullying detection in OSNs~\cite{Improved} . Huang  {\em et al.} \cite{huang2014cyber} has consider some graph properties besides text features, however they also worked only over comment-based labeled data. Another work has looked at the time series of posted comments of Formespring dataset, which each question answer pair was labeled separately as cyberaggression and predict their severity, \cite{potha2014cyberbullying}. Also much of these works have considered text-based approach to analyze comments in OSNs instead of analyzing the features associated with the media objects such as images or videos belonging to those comments.  Kansara  {\em et al.} \cite{kansara2015framework} suggest only a framework for using images beside text for detecting cyberbullying. 

Other work analyzed profanity usages in Instagram \cite{homa-cyberbullying-socialcom14} and Ask.fm \cite{asonam14} comments, but did not label the data at all. Our previous work, \cite{HosseinmardiMRH15} detects cyberbullyning incidents in the Instagram for images with comments with more than 40\% negativity. Additional research investigated aspects of the Instagram social network, but not in the context of cyberbullying.  For example, \cite{Weilenmann} explored users' photo sharing experience in a museum.  Silva {\em et al.} \cite{insta} considered the temporal photo sharing behavior of Instagram users and  Hu  {\em et al.} \cite{whatinsta} categorized Instagram images into eight popular image categories and the Instagram users into five types in terms of their posted images. By investigating user practices in Instagram, \cite{araujo2014} concluded that users tend to be more active during weekends and at the end of the day. They also found out that users are more likely to like and comment on the medias that are already popular, thereby inducing the rich get richer phenomenon.

\section{Data Collection}
Starting from a random seed node, we identified 41K Instagram user ids using a snowball sampling method from the Instagram API.  Among these Instagram ids, 25K (61\%) users had public profiles while the rest had private profiles. 
Due to the limitation on the private profiles' lack of shared information, the 25K public user profiles comprise our sample data set. For each public Instagram user, the collected profile data includes the media objects (videos/images) that the user has posted and their associated comments, user id of each user followed by this user, user id of each user who follows this user, and user id of each user who commented on or liked the media objects shared by the user. We consider each media object plus 
its associated comments as a {\em media session}. 

Labeling data is a costly process and therefore in order to make the labeling of cyberbullying more manageable, we sought to label a smaller subset of these media sessions. To have a higher rate of cyberbullying instances, we considered media sessions with at least one profanity word in their associated comments. We tag a comment as ``negative" using an approach similar to \cite{asonam14}. For this set of 25K users, 3,165K unique media sessions were collected, where 697K of these sessions have at least one profane word in their comments by users other than the profile owner, where a profane word is obtained from a dictionary \cite{noswearing}, \cite{NegativeWordsList}. 

In addition, we needed media sessions with enough comments so that labelers could adequately assess the frequency or repetition of aggression, which is an important part of the cyberbullying definition.  We selected a threshold of 15 as a lower bound on the number of comments in a media session, considering that the average ratio of comments posted by users other than friends to comments posted by the profile owner in an Instagram profile is around 16 \cite{homa-cyberbullying-socialcom14}. At the end 2,218 media sessions (images and their associated comments) were selected randomly for the task of labeling. 

\section{Cyberbullying Labeling}
\label{labeling}
In this section, we explain the design and methodology for our survey for labeling the selected set of media sessions. Our first challenge is choosing appropriate definitions of terms, which will then be used in ground truth labeling.  Based on the literature, a major early choice that we have made is to distinguish between cyberaggression and cyberbullying.  Cyberaggression is broadly defined as any occurrence of using digital media to intentionally harm another person~\cite{kowalski2012cyberbullying} .  Examples include negative content and words such as profanity, slang and abbreviations that would be used in negative posts such as hate, fight, wtf.
Cyberbullying is one form of cyberaggression that is more restrictively defined as intentional aggression that is repeatedly carried out in an electronic context against a person who cannot easily defend him or herself \cite{kowalski2012cyberbullying,patchin2012update}. 
Thus, cyberbullying  consists of three main features : (1) an act of aggression online; (2) an imbalance of power between the individuals involved; and (3) it is repeated over time \cite{hunter2007perceptions,kowalski2012cyberbullying,olweus1993bullying,olweus2013school,Smith2012}. The power imbalance can take on a variety of forms including physical, social, relational or psychological \cite{dooley2009cyberbullying,monks2006definitions,olweus2013school,pyzalski2010electronic}, such as a user being more technologically savvy than another~\cite{kowalski2014bullying}, a group of users targeting one user, or
a popular user targeting a less popular one \cite{Limber2008}. Repetition of cyberbullying can occur over time or by forwarding/sharing a negative comment or photo with multiple individuals \cite{Limber2008}. 

In Instagram, each media session consists of a media object posted by the profile owner and the corresponding comments for the media object. The goal in this paper is to investigate cyberaggression and cyberbullying in this {\it multi-modal context} (textual comments and media objects).
Our labeling process consisted of two separate surveys. The first survey
incorporates both the image and the associated text comments when asking the human contributors whether the media session was an instance of cyberbullying or cyberaggression. The second survey is focused on the image contents alone and is used
to identify the content and category of the image. 

\begin{figure}[!ht]
\centering
\includegraphics[width=0.5\textwidth]{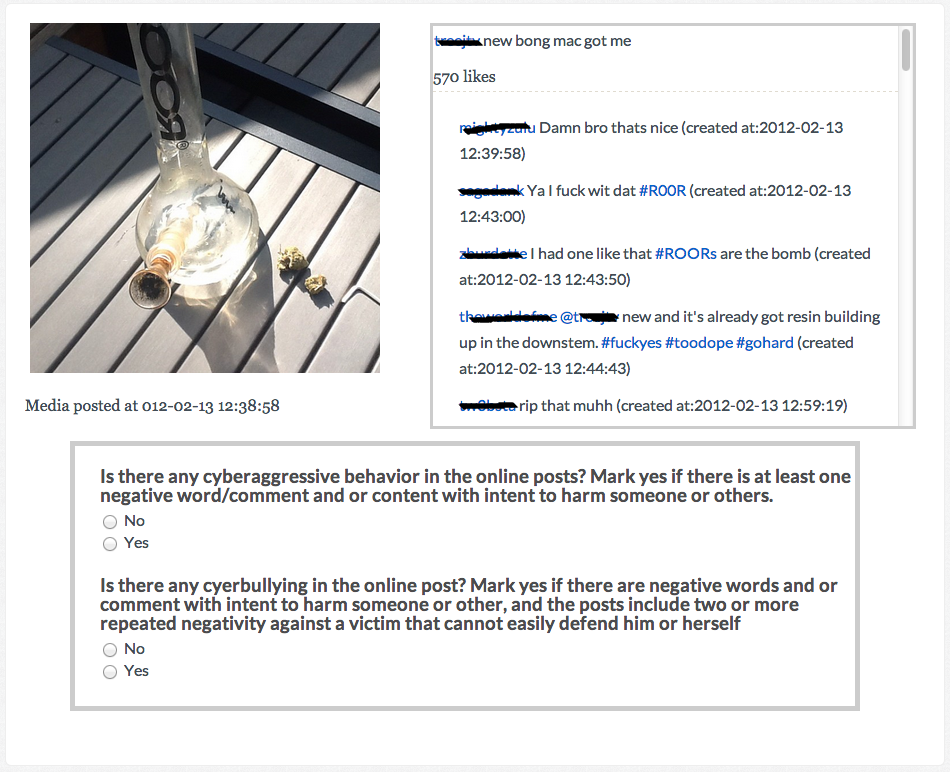}
\vspace{-2.0em} 
\caption{An example of the labeling survey, showing an image and its corresponding comments, and the survey questions.}
\label{fig:survey}
\end{figure}

Figure~\ref{fig:survey} illustrates an example of our design for the first labeling survey.  On the left is the image, and on the right is a scrollable interface to help the contributor see all of the comments associated with this image. 
With input from a social science expert, 
we designed a simple survey that asked the contributors two questions, namely whether the media session constituted cyberaggression or not, and whether the media session constituted cyberbullying or not.  During the instructional phase prior to labeling, contributors were given the aforementioned definitions of cyberaggression and cyberbullying along with related examples.  Each media session was labeled by five contributors.

\subsection{Quality Control}
In order to provide quality control, we only permitted the highest-rated contributors on CrowdFlower to have access to our job. Next, a mentoring phase was provided for the potential contributors that included instructions and a set of example media sessions with the correct label. Further, to monitor the quality of the contributors and filter out the spammers, potential contributors were asked to answer a set of test questions in two phases: quiz mode and work mode. Potential contributors needed to answer correctly a minimum number of test questions to pass the quiz mode and qualify as a contributor for the survey. 
We also incorporated quality control checks during the labeling process (work mode) by inserting random test questions. A contributor was filtered out if he/she failed this work mode. Finally, a minimum time threshold was set to filter out contributors who rushed too quickly through the labeling process. The minimum number of test questions to get back high-quality data was recommended by CrowdFlower.

Overall, 176 potential contributors worked on the quiz questions, 144 passed the quiz mode, while 31 contributors failed and 1 gave up. The labeled data that we finally obtained were from 139 {\it trusted} contributors, while the the rest were filtered out during the work mode. Table \ref{labeling:stats1} provides the number of trusted judgments and the contributors' accuracy for 11,090 total judgments.

\begin{table}[htbp!]
\centering

\caption{Labeling process statistics. Trusted judgments are the ones made by trusted contributors.}
\vspace{1.5mm}
\begin{tabular}{ p{5cm}  c  }
\hline 
    \hline
   Trusted Judgments  & 10987  \\  \hline
   Untrusted Judgments & 103 \\ \hline
   Average Test Question Accuracy of Trusted Contributors &	89\% \\ \hline
   Labeled Media Sessions per Hour & 6 \\ \hline
    \hline
\end{tabular}

\label{labeling:stats1}
\end{table}

\subsection{Image Labeling}
Next, we conducted a second separate survey just to label the image contents of media sessions, so that we could investigate the relationship between cyberbullying and image content.  
We first sampled 1,200 images from the selected subset of media sessions to determine a suitable set of representative categories to be used in the labeling. A graduate student examined all the images and classified them to different possible categories. Then for designing the survey, a social science expert checked the categories again and revised them. Some of the dominant categories identified were the presence of a human in the image, as well as text, clothes, tattoos, sports and celebrities. We then 
asked contributors to identify which of the aforementioned categories were present in the image.  Multiple categories could be selected for a given image. Each media session was labeled by three different contributors.

At the end, our social science expert checked a set of random media sessions and images to confirm the quality of the labeled data for both surveys.


\section{Analysis and Characterization of Ground Truth Data}
\label{sec:analysis}

We submitted our first survey with 2,218 media sessions (images and their associated comments) to CrowdFlower, a crowd-sourced website. Each media session was labeled by five different contributors.  CrowdFlower provides a degree of trust for each contributor based on the percentage of correctly answered test questions explain in section \ref{labeling}.  This trust value is incorporated by CrowdFlower into a weighted version of the majority voting method called a ``confidence level" for each labeled media session.  We decided to keep media sessions whose weighted trust-based metric was equal to or greater than 60\%. We deemed them to be strong enough support for majority voting from contributor with higher trust.  Overall, 1,954 (88\%)  of the original pure majority-vote based media sessions wound up in this higher-confidence cyberbullying-labeled group.  For this higher-confidence data set, 29\% of the media sessions belonged to the ``bullying" group while the other 71\% were deemed to be not bullying. 

\subsection{Labeling and Negativity Analysis}
\begin{figure}[lht]
\centering
\includegraphics[width=0.5\textwidth]{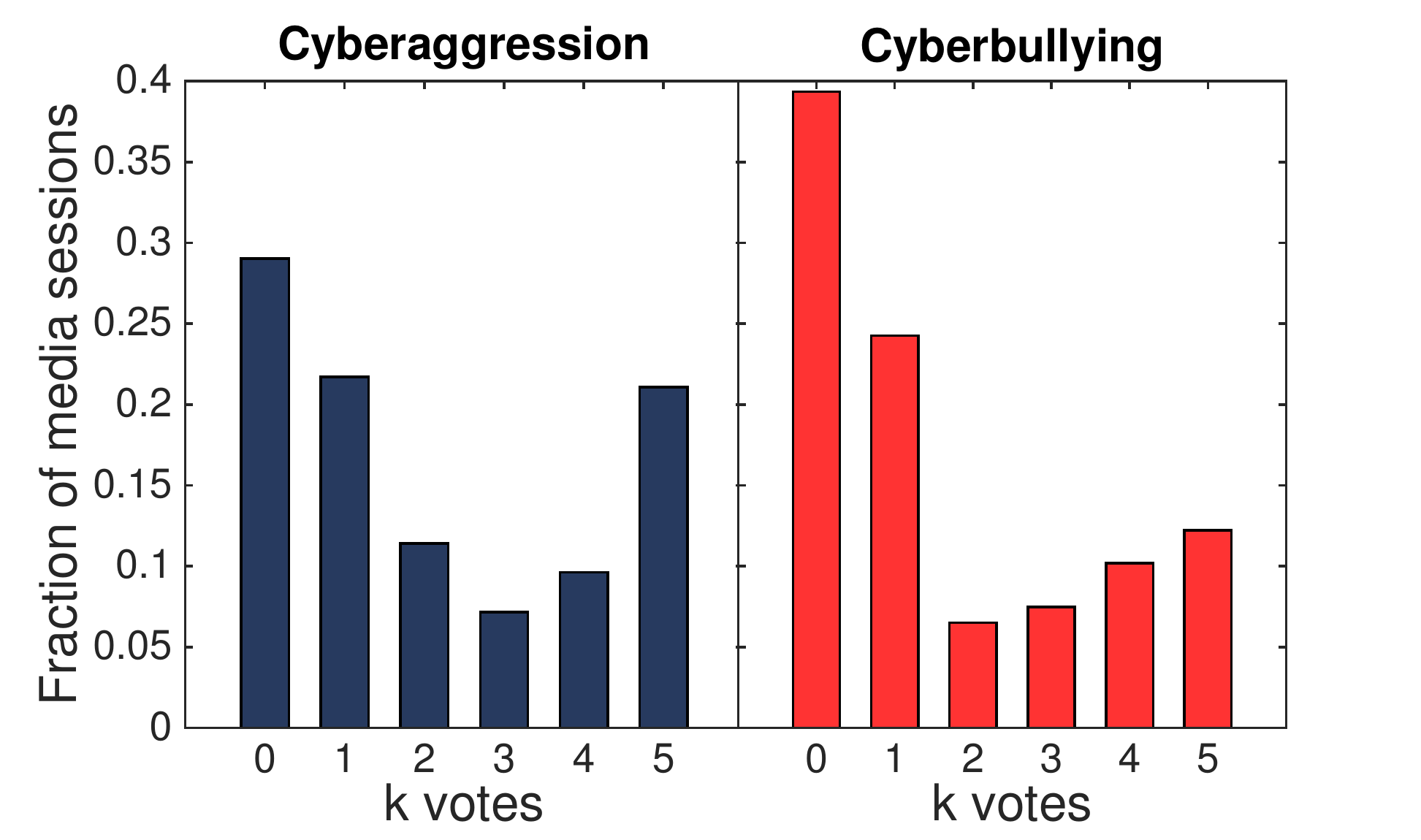}
\vspace{-1.0em}
\caption{Fraction of media sessions that have been voted $k$ times as cyberaggression (left) or cyberbullying (right).}
\label{fig:dist_votes}
\end{figure}

 Distribution of the media sessions based on the number of votes (out of five votes) received for cyberaggression and cyberbullying respectively has been provided in Figure~\ref{fig:dist_votes}. 
The left chart shows the fraction of samples that have been labeled as cyberaggression $k$ times, and the right chart shows the fraction of samples that have been labeled as cyberbullying $k$ times.  
Higher the number of votes for a given media session, more confidence we have that the media session contains an incident of cyberaggression or cyberbullying, with five votes means unanimous agreement.
Similarly, lower the number of votes for a given media session, more confidence we have that the media session {\it does not} contain an incident of cyberaggression or cyberbullying, with zero votes means unanimous agreement.The inter-rater agreement Fleiss-Kappa value for cyberbullying is 0.5 and for  cyberaggression is 0.52.

We notice that for both cyberaggression and cyberbullying, most of the
probability mass is around media sessions labeled by all four or five contributors the
same, i.e. either 0 or 1 votes (about 50\% for cyberaggression and about 62\% 
for cyberbullying), or 4 or 5 votes (about 31\% for cyberaggression and
about 23\% for cyberbullying).
\emph{Thus, a key finding is that the contributors are mostly in agreement about what behavior constitutes cyberaggression, and what behavior constitutes cyberbullying in Instagram media sessions.}  
Only about 13$-$17$  $\% of the media sessions have two or three votes, which 
indicates that there is some disagreement in a small fraction of media
sessions about whether the session contains an incident of cyberaggression or
 cyberbullying.
This disagreement can be attributed to the fact that different people have different levels of sensitivity and a conversation may seem normal to one person and hurtful to another. 

Next, we observe that about 30\% of the media sessions have not been labeled as cyberaggression by any of the five contributors. Since all media sessions contained at least one comment with one or more profane word, this suggests that only
employing a profanity usage threshold to detect cyberaggression can produce
many false positives.
We make a similar observation for cyberbullying. We notice about 40\% of the media sessions have not been labeled as cyberbullying by any of the
five contributors. Applying a majority voting criterion to a binary label as cyberbullying or not, 30\% of the samples have been labeled as cyberbullying.  This is despite the fact that all the media sessions contain at least one profane word. This leads us to our second important finding.
{\it A classifier design for cyberbullying detection cannot solely rely on the usage of profanity among the words in image-based discussions, and instead must consider other features to improve accuracy.}

\begin{figure}[!ht]
\centering
\includegraphics[width=0.35\textwidth]{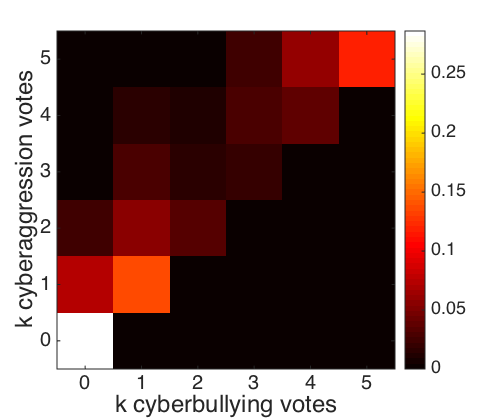}
\caption{Two-dimensional distribution of media sessions as a function of the number of votes given for cyberaggression versus the number of votes given for cyberbullying, assuming five labelers.}
\label{fig:heat1}
\end{figure}

In order to understand the relationship between cyberaggression and cyberbullying, we plotted in Figure~\ref{fig:heat1} a two-dimensional heat map that shows the distribution of media sessions as a function of the number of votes each media session received for cyberaggression and cyberbullying.  We observe that a significant fraction of the sessions exhibit strong agreement in terms of either receiving high numbers of votes for both cyberbullying and cyberaggressions, or receiving low numbers of votes for both cyberbullying and cyberaggression. This can be inferred from the high energy in the upper right and lower left part of the diagonal. In addition, it is promising that the area below the diagonal is essentially zero, meaning no session has received more votes for cyberbullying than for  cyberaggression.  This conforms with the definition that cyberbullying is a subset of cyberaggression. The Pearson's correlation between number of vote for cyberbullying and number of votes for cyberaggression is 0.9.  

We see that the remaining significant energy in the distribution appears in the area above the diagonal.  Media sessions in this area exhibit the property that if they receive $N_1$ cyberbullying votes and $N_2$ cyberaggression votes, then $N_2 \geq N_1$. The area where $N_1\leq 2$ and $N_2 \geq 3$ corresponds to cases where there is cyberaggression but not cyberbullying. These observations lead us to our third important finding. \emph{A media session that exhibits cyberaggression does
not necessarily exhibit cyberbullying, and a classifier design for cyberbullying detection must go much beyond merely detecting cyberaggression.} This is a
very important finding, because as we noted in Section \ref{sec:related}, prior work on detecting cyberbullying has mainly focused on detecting cyberaggression as they do not take into account the frequency of aggression or imbalance of power, which are crucial features of cyberbullying.

\begin{figure}[!ht]
\centering
\includegraphics[width=0.5\textwidth]{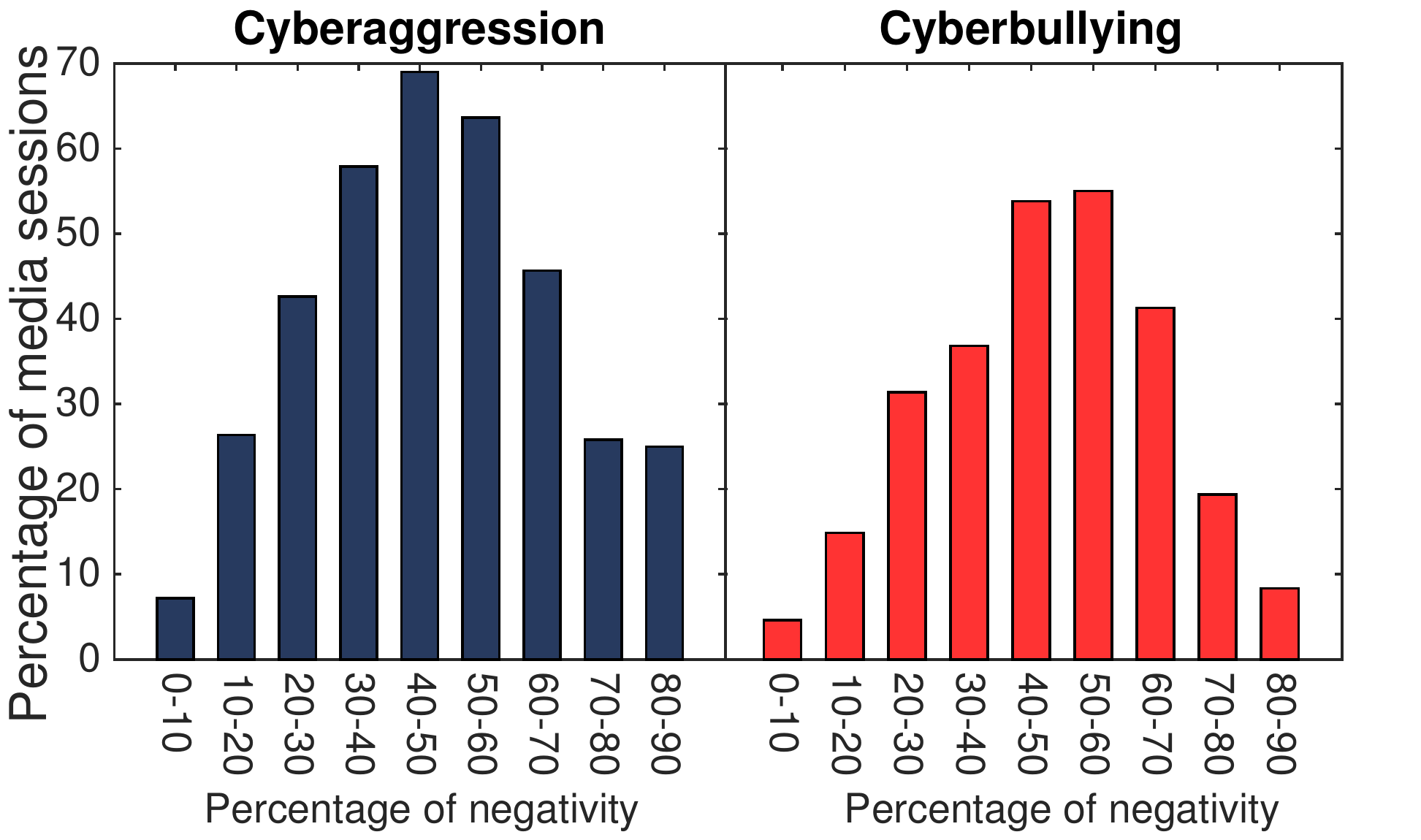}
\caption{Percentage of media sessions that have been labeled as cyberaggression (left) and cyberbullying (right) versus their negativity percentages.}
\label{fig:lblvsngtvty}
\end{figure}

Finally, we are interested in understanding the relation between cyberbullying/cyberaggression and the percentage of negativity in the comments. We divided all the media sessions into nine different bins based on the percentage of negativity in their comments. Bin $(n_1-n_2]$ contains all media sessions with bigger than $n_1\%$ and smaller than or equal to  $n_2\%$ negativity. None of the media sessions contained more than 90\% negative comments. 
Next, we calculated percentage media sessions for each bin that can be identified as cyberaggression or cyberbullying based on majority of votes, i.e. where the number of votes is 3 or higher.
Figure \ref{fig:lblvsngtvty} shows these fractions, left figure for cyberaggression and right figure for cyberbullying. We observe that as the percentage of negativity increases, so does the fraction
of media sessions up until 50\% negativity for cyberaggression and 60\% for
cyberbullying. This increase is as expected, since cyberaggression or
cyberbullying is typically accompanied with negativity in the postings.
However, we notice that the percentage of cyberaggression or cyberbullying
starts decreasing after these peaks as the percentage of negativity increases.
This is quite an unexpected result and seems counter-intuitive.
To understand this, we examined closely the media sessions that have very
high negativity. We noticed that these media sessions typically involved 
discussions about sports, politics, tattoos, or were just friendly talks. 
People tend to use lots of profanity words in such discussion, even though
they are not insulting any one person in particular.
This leads us to our final important finding about negativity analysis.
\emph{A media sessions with a significantly high percentage of negativity
(more than 60-70\%) typically implies a low probability that the session contains a cyberbullying incident.}

\subsection{Temporal and Graph Properties Analysis}

Since different comments in a media session are posted at different times, 
it is important to understand the relationship between the temporal nature
of comment postings and cyberbullying/cyberaggression.  
We define the strength of cyberbullying/cyberaggression as the number of votes received for labeling a media session as cyberbullying/cyber-aggression,
and explore the Pearson's correlation of cyberbullying/cyberaggression strength and temporal behavior comment arrivals. We would like to understand how human contributors incorporated the definition of cyberbullying, which includes the temporal notion of repetition of negativity over time, into their labeling.  Given the time stamps on the collected comment, we compute the interarrival time between two consequent comments.  We then count the number of interarrival times of comments in a media session that have a value less than $x $ = 1min, 5 min, ...,6 months.

\begin{figure}[!ht]
\centering
\includegraphics[width=0.47\textwidth]{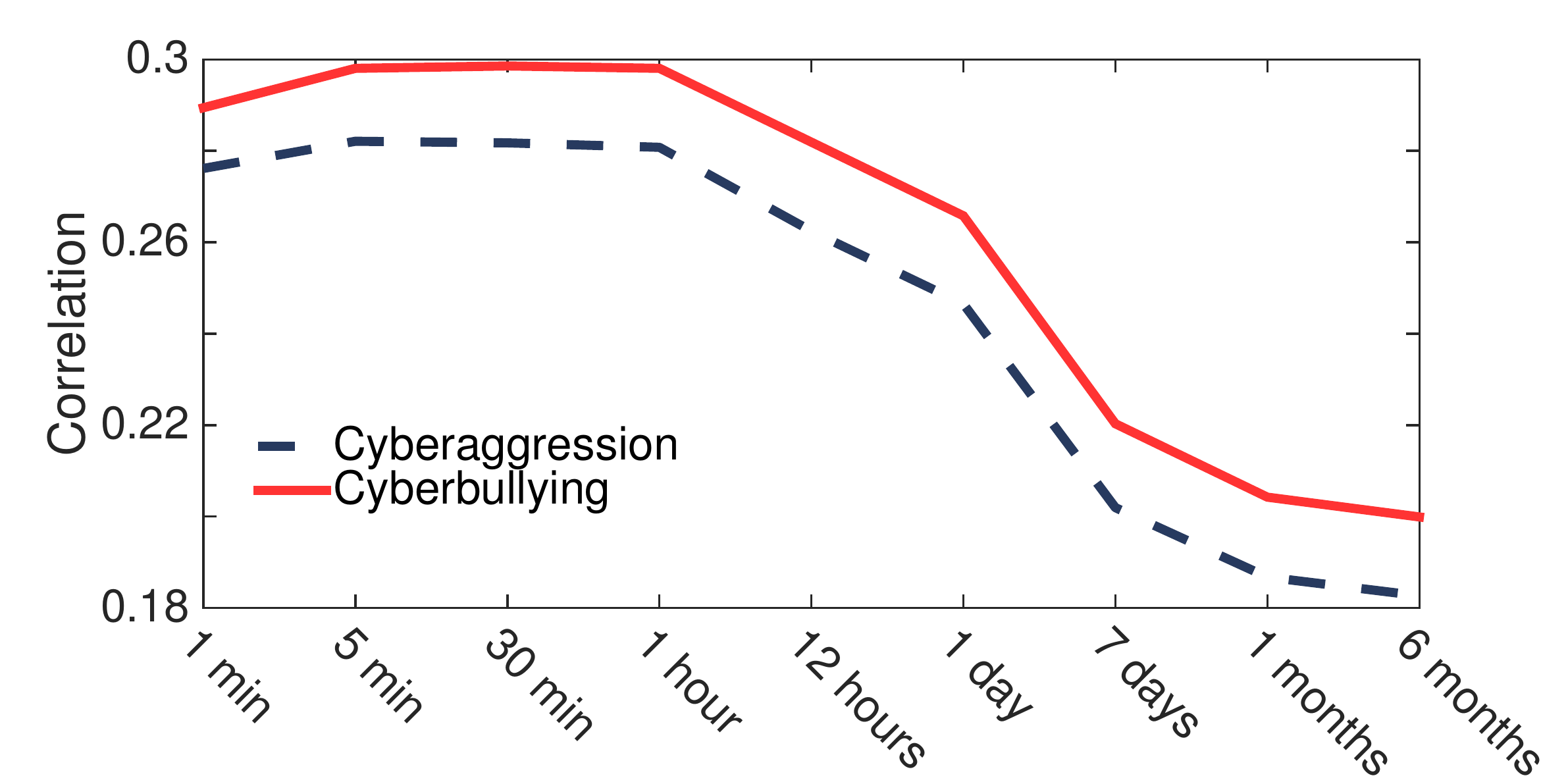}
\vspace{-4mm}
\caption{Pearson's correlation between the number of votes and the number of comments arrive with less than or equal $x$ seconds after their previously received comment. (X-axis has been converted to coarser time units than second)}
\label{fig:temporal}
\end{figure}

Figure~\ref{fig:temporal} illustrates the correlation between the number votes and the number of comments arrive with less than or equal $x$ seconds after their previously received comment. We see that there is a
correlation of about 0.3 between the strength of support for cyberbullying and media sessions in which there are frequent postings within one hour of previous post.  Further, we find that as we expand the allowable interarrival times between comments, the correlation weakens considerably. Similar pattern was observed for cyberaggression. In fact, on average around 40\%  of the comments arrive in less than 1 hours after previously received comment in cyberbullying media sessions, however only 30\% of the comments have been received with the same interarrival time in non-cyberbullying samples ($p < 0.001$, based on t-test). 
{\it A key finding here is that media sessions that contain cyberbullying
have relatively low comment interarrival times, that is the comments in 
these media sessions are posted quite frequently.}
 
 \begin{table*}[htbp!]
\centering
\caption{Mean values of social graph properties for cyberbullying versus non-cyberbullying samples and aggression versus non-cyberaggression. $(*p<0.05)$.}
\vspace{2mm}
\begin{tabular}{ l  c  c  c  c  }
\hline 
    \hline
    Label & *Likes & *Media objects & Following & Followers \\ \hline
   Non-cyberbullying  & 9,684.4 &  1,145.7   & 668.1 &  415,676.2  \\ \hline
    Cyberbullying  &  7,029.0 &  1,198.3    & 626.7  &  463,073.1  \\ \hline
       Non-cyberaggression & 9,768.6  & 1,133.7 &   665.9    &  421,075.3  \\ \hline
    Cyberaggression  &  7,551.3  &   1,204.3  & 640.3  &   440,403.6  \\ \hline
    \hline
\end{tabular}

\label{graph_stats1}
\end{table*}

Next, we examine the relationship between cyberbullying/cyberaggression
and the social network graph features such as the number of likes for a
given media object, number of comments posted for a media object, 
number of users a user is following, and the number
of followers of a user.
Table~\ref{graph_stats1} shows these numbers, for categories of non-cyberbullying
sessions, cyberbullying sessions, non-cyberaggression sessions and cyberaggression sessions.
We observe that media sessions that contain cyberbullying/cyberaggression share more media objects than media sessions that do not contain cyberbullying/cyberaggression, but on average receive lower number of likes. Souza {\em et al.}'s \cite{araujo2014} analysis of Instagram users shows there is a positive correlation between number of followers and number of likes for typical Instagram users. Users who receive cyberbullying do not follow the same pattern. In fact, the average number of likes per post for non-cyberbullying sessions is 4 times the average number of likes for cyberbullying sessions, and the the average number of likes per post for 
non-cyberaggression sessions is 4.5 times the average number of likes for cyberaggression sessions. In terms of number of following and followers, the
distinction is not as pronounced, although we see that the media sessions with cyberbullying/cyberaggression incidents have more
followers and less following compared to the media sessions without cyberaggression/cyberbullying. {\it The key finding here is that
the users of media sessions with cyberbullying/cyberaggression have lower number of likes per post while have more followers.}

\subsection{Linguistic and Psychological Analysis}
\label{sec:ling}
\begin{figure}[!ht]
\centering
\includegraphics[width=0.51\textwidth]{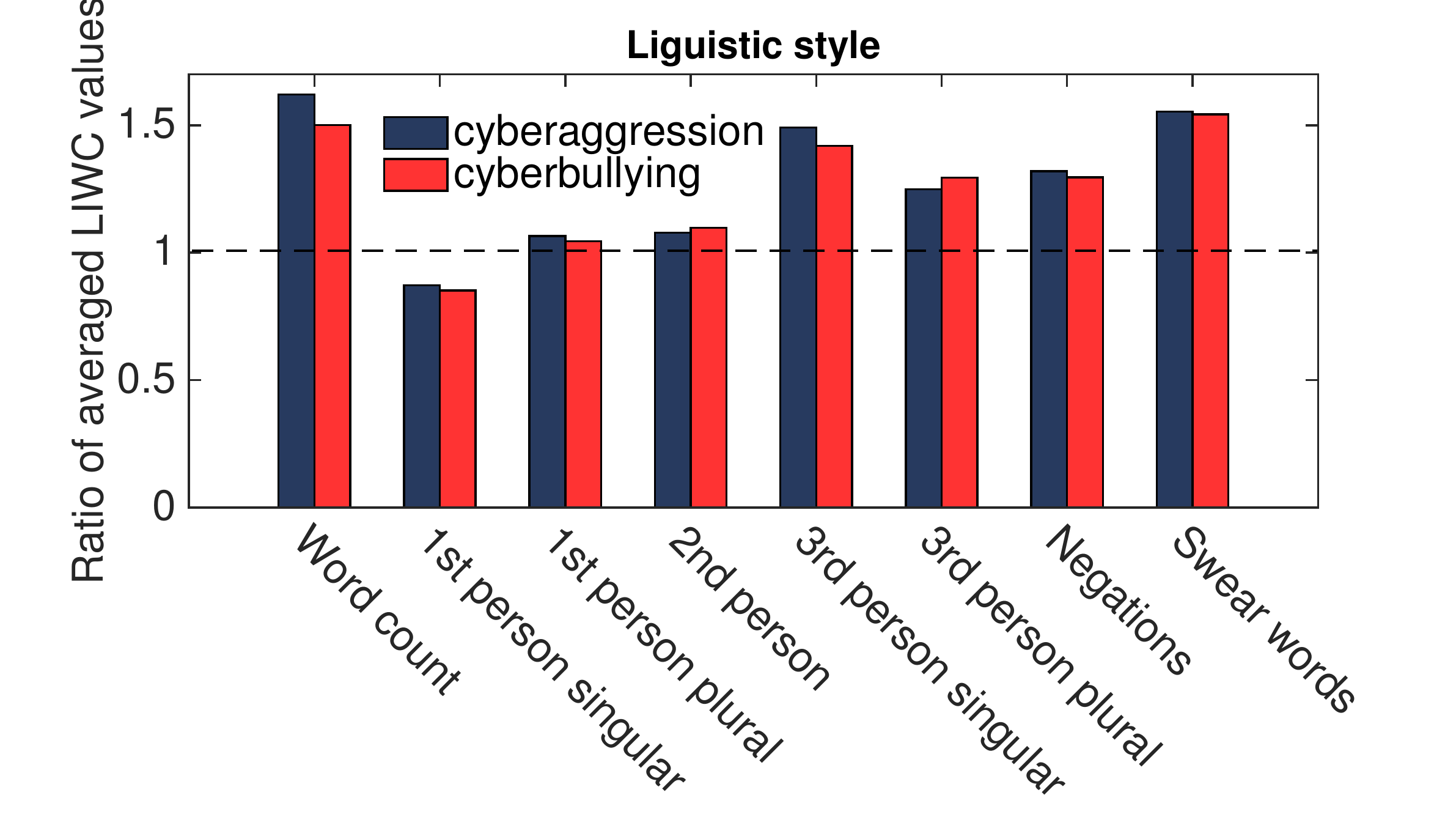}
\vspace{-8mm}
\caption{Ratio of LIWC values of cyberbullying/cyberaggression labeled media sessions to non-cyberbullying/non-cyberaggression class in Linguistic categories.}
\label{fig:liwc3}
\end{figure}

\begin{figure}[!ht]
\centering
\includegraphics[width=0.51\textwidth]{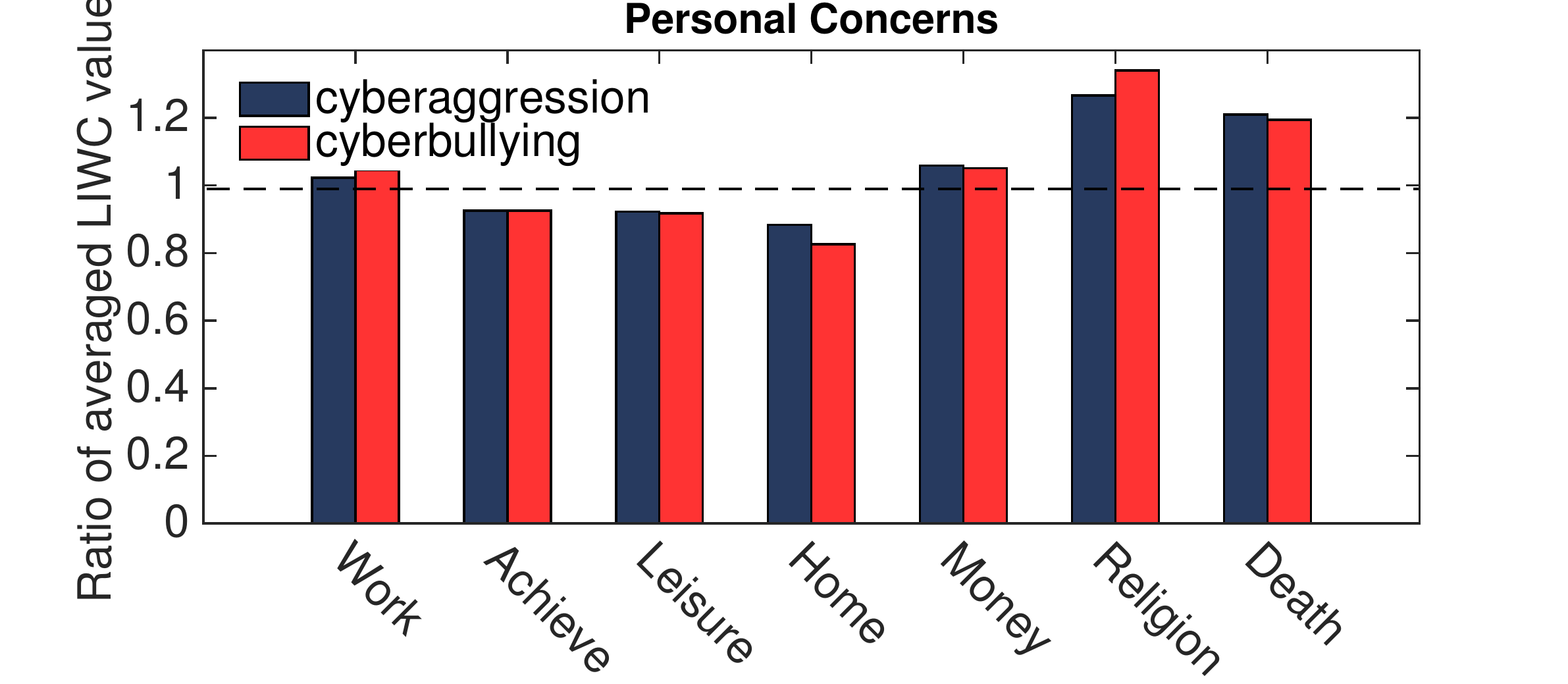}
\vspace{-6mm}
\caption{Ratio of LIWC values of cyberbullying/cyberaggression labeled media sessions to non-cyberbullying/non-cyberaggression class in Personal Concerns categories.}
\label{fig:liwc2}
\end{figure}

\begin{figure*}[!ht]
\centering
\includegraphics[width=0.8\textwidth]{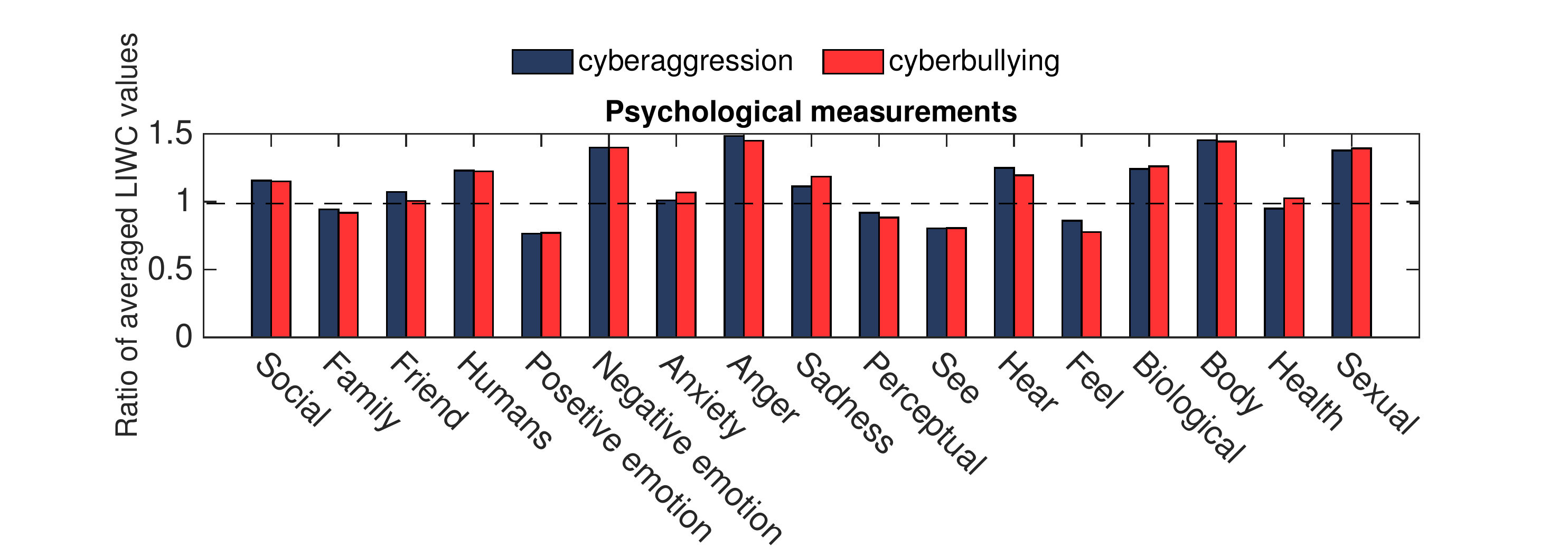}
\vspace{-3mm}
\caption{Ratio of LIWC values of cyberbullying/cyberaggression labeled media sessions to non-cyberbullying/non-cyberaggression class in Psychological categories.}
\label{fig:liwc1}
\end{figure*}

We now focus on the pattern of linguistic and psychological measurements of cyberbullying/cyberaggression media sessions versus non-cyberbullying/non-cyberaggression. For this 
purpose,  we have applied Linguistic Inquiry and Word Count (LIWC) a text analysis program to find which categories of words have been used for cyberbullying/cyberaggre-ssion labeled media sessions. LIWC evaluates different aspects of word usages in psychologically meaningful categories, by counting the number of the words across the text for each category \cite{liwc}.  LIWC has often been used for studies on variations in language use across different people. Published papers show that LIWC have been validated to perform well in studies on variations in language use across different peoples~\cite{liwc1}. We first analyze the number of words, and usage of pronouns, negations and swear words 
(Figure~\ref{fig:liwc3}). Next, we look at some of the personal concerns
such as work, achievements, leisure, etc. (Figure~\ref{fig:liwc2}). Finally,
we investigate some of the psychological measurements such as social, family, 
friends, etc. (Figure~\ref{fig:liwc1}).
For each of these cases, we first obtain the LIWC values for each media
session comment set. We then calculate the average LIWC value for each of
the four classes: 
media sessions with cyberbullying, media sessions with no cyberbullying,
media sessions with cyberaggression, and media sessions with no
cyberaggression. The bars shown in Figures~\ref{fig:liwc3}-\ref{fig:liwc1}
represent the ratio of average LIWC value for cyberbullying class to
that of non-cyberbullying, and the ratio of average LIWC value for
cyberaggression class to that of non-cyberaggression.

In Figure \ref{fig:liwc3}, we first notice that the word count for
media sessions with cyberbullying/cyberaggression is significantly higher
than for media sessions with no cyberbullying/cyberaggression ($p< 10^{-5}$). Next,
as expected, for swear words (e.g., damn, piss) and negations (e.g., never,
not), the ratio is higher for cyberbullying/cyberaggression 
category ($p< 10^{-5}$).
It is interesting to note that the ratios for the third person pronouns
(she, he, they) are more than 1.3 ($p< 10^{-5}$), ratio
for the first person singular pronoun ( i ) is 0.85, and the ratios for
first person plural and second person pronouns (we, you)
is closer to 1. This leads us to our first key finding with respect to
the linguistic features. 
{\it A user is less likely to directly refer to himself/herself and more
likely to refer to other people in third person in postings involving
cyberbullying or cyberaggression.}

For personal concerns (Figure \ref{fig:liwc2}), ``religion" (e.g., church, mosque) and ``death" (e.g., bury, kill) categories have higher ratios (more than
1.2, p $<0.1$). This is in line with our findings in our previous work on
profanity usage analysis in ask.fm social media, where we observed that
there is high profanity usage around words like ``muslim" \cite{asonam14}.
This confirms that religion-based cyberbullying is quite prevalent in
social media. On the other hand, ratios for personal categories like ``work",
``money" and `achieve" are much closer to 1. 

For psychological measurements (Figure \ref{fig:liwc1}), we notice that the
ratios for ``negative emotion'', ``anger'', ``body'',
and ``sexual'' categories are
significantly higher than 1 (more than 1.4, $p< 10^{-5}$), and the ratio for
``positive emotion''
category is significantly lower than 1 (0.76, $p< 10^{-5}$).
Higher ratios for ``body'' (e.g. face, wear) and ``sexual'' (e.g.
slut, rapist) categories provide
evidence for appearance-based and sexual-based cyberbullying in social
media. For other psychological measurement categories, such as
``social'', ``friend'', etc., the ratios are closer to 1.
Based on our observations from Figures~\ref{fig:liwc2} and \ref{fig:liwc1},
our final important finding with respect to linguistic
features is as follows: {\it There is a higher probability of cyberbullying
in postings involving religion, death, appearance and sexual hints, and
cyberbullying posts typically have higher occurrences of negative emotions
and lower occurrences of positive emotions.}

\subsection{Image Content Analysis}

\begin{figure}[!ht]
\centering
\includegraphics[width=0.5\textwidth]{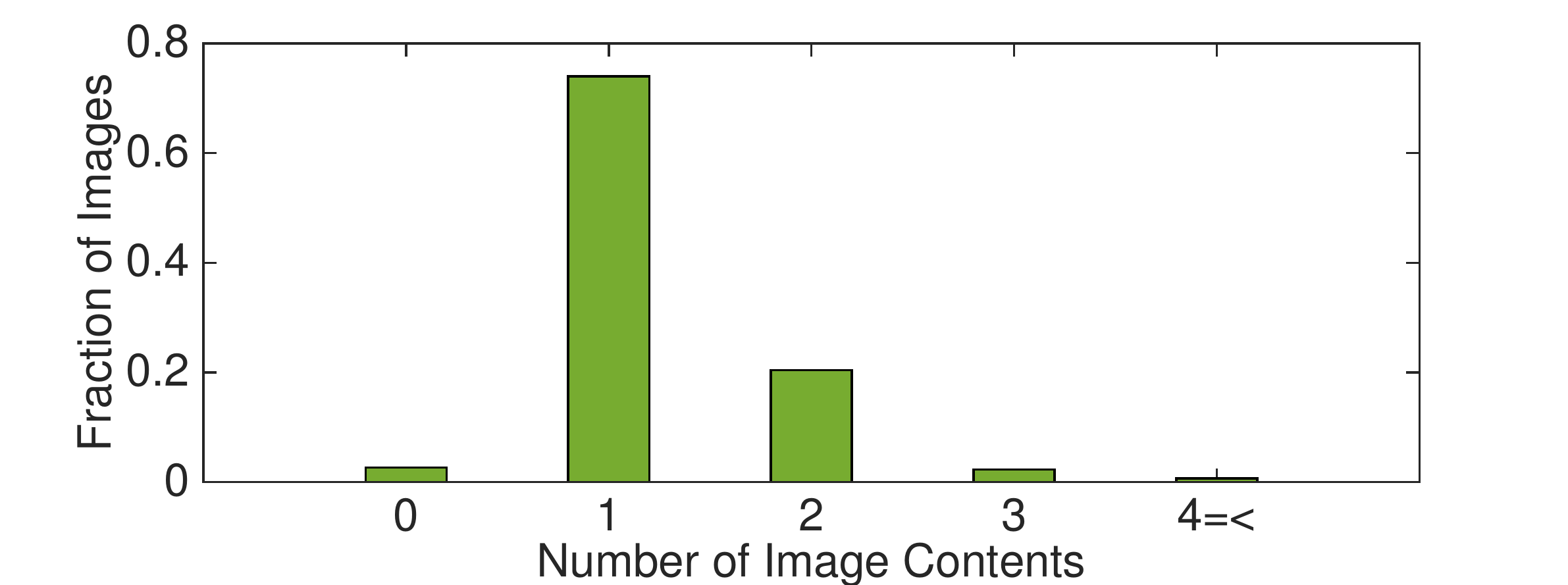}
\vspace{-2.0em} 
\caption{Distribution of number of unique contents assigned to each image in multichoice labeling.}
\label{fig:imagemultichoice}
\end{figure}

We now explore the relationship between image content and cyberbullying/cyberaggression in a media session.  Recall that contributors could place an image in more than one category. More than 70\% of the images were labeled with only one category, and around 20\% were labeled with two categories. However, there were few images that were labeled with up to eight unique categories. Figure \ref{fig:imagemultichoice} shows the distribution of the number of contents which have been assigned to the images by CrowdFlower contributors.
We used majority mechanism to decide on a single content category for each image. In case of a tie, a graduate student looked at those images to break
the tie.

\begin{figure*}[!ht]
\centering
\includegraphics[width=0.8\textwidth]{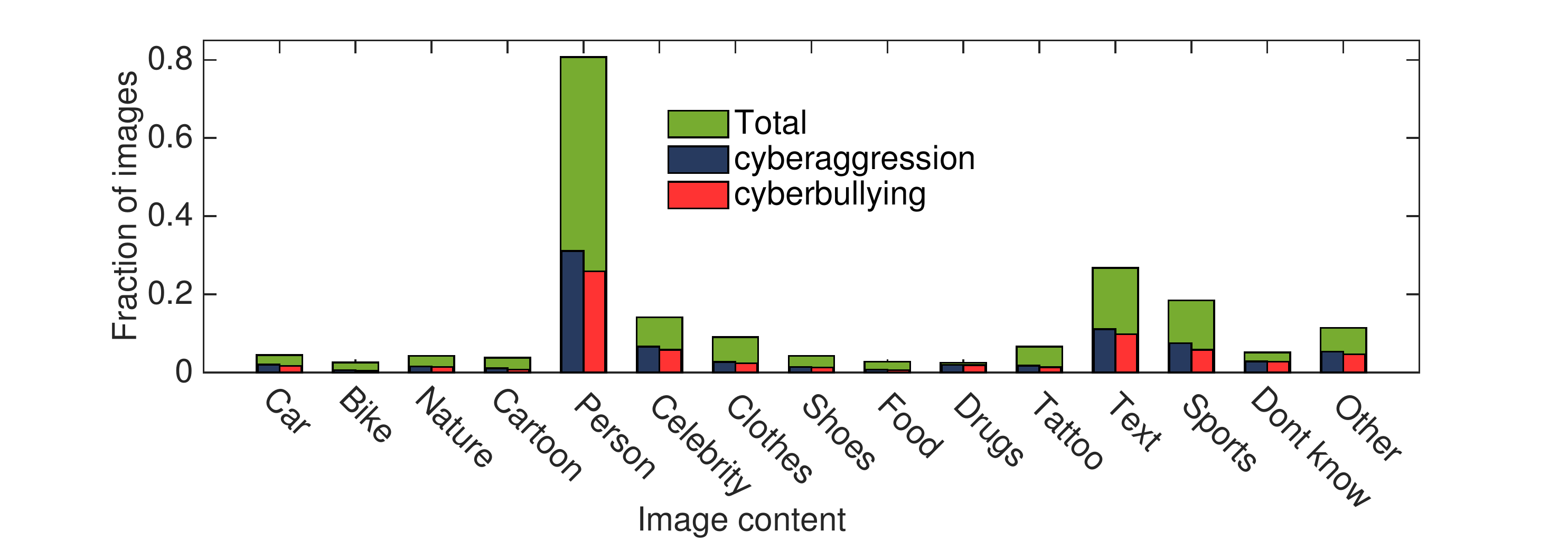}
\vspace{-3mm}
\caption{Fraction of image categories for all media sessions, cyberbullying and cyberaggression classes.}
\label{fig:image_dist}
\end{figure*}

Figure \ref{fig:image_dist} shows the fraction of the contents for all labeled data in green bar.
The blue and red bars embedded inside each green bar in this figure indicate
the fraction of images that belonged to the media sessions that can be 
categorized as containing cyberaggression and cyberbullying respectively.
The ``dont know" choice was given as we realized that for some images it is hard to figure out what is in the image. As some images belong to more than one category, the bars will sum up to more than one. 

\begin{figure*}[!ht]
\centering
\includegraphics[width=0.8\textwidth]{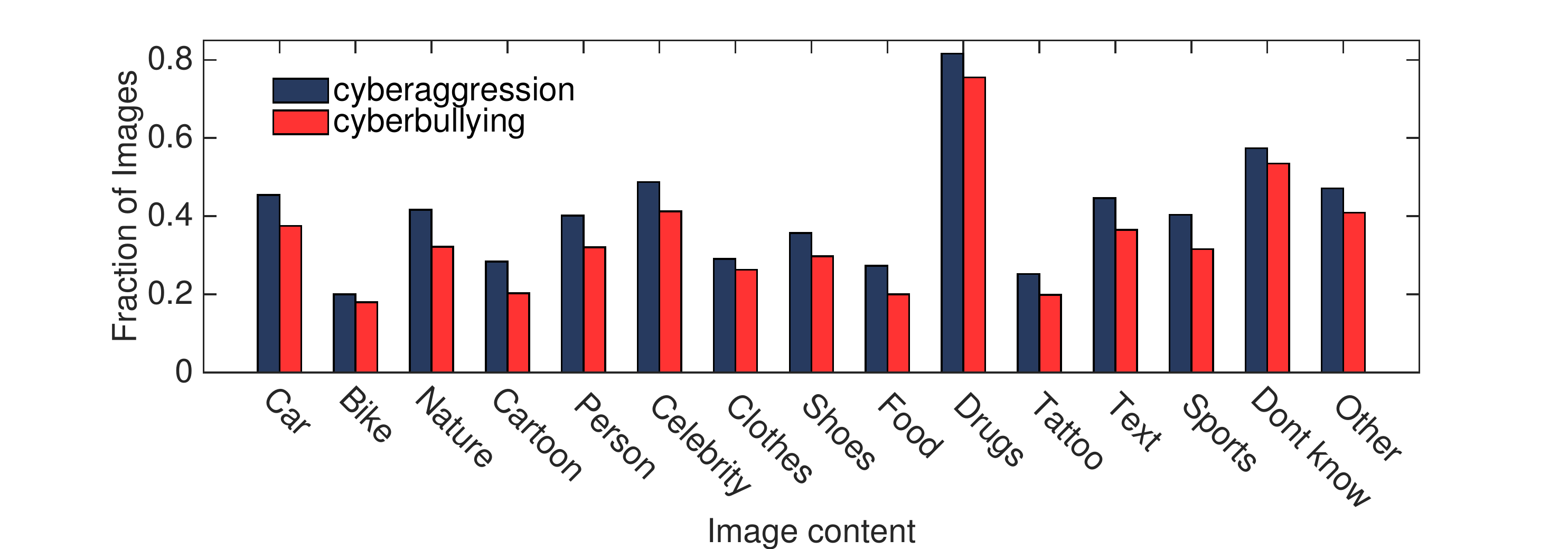}
\vspace{-3mm}
\caption{Fraction of images which have been labeled as cyberbullying and cyberaggression  for each content category.}
\label{fig:imagemultichoiceratio}
\end{figure*}

First, we observe that the most common labels for image content are Person/People, Text and Sports.
Second, we observe that for some content categories such as ``Drugs'', the
fraction is quite small, but most of the images in those categories do
belong to media sessions with cyberaggression/cyberbullying in them. To understand this further,
in Figure~\ref{fig:imagemultichoiceratio}, we show the fraction of images labeled as cyberaggression/cyberbullying
for each content category.
We notice that for content category ``Drugs'', 75\% of the images belong
to media sessions containing cyberbullying, while  for content categories
like ``Car'', ``Nature'', ``Person'', ``Celebrity'', ``Text'' and ``Sport'',
30\%-40\% of the images belong 
to media sessions containing cyberbullying/cyberaggression.
Also, whenever images contain bike, food, tattoo, etc., there is 
little cyberbullying occurring. 
{\it The key finding here is that certain image contents such as Drug are
highly correlated with cyberbulllying, while some other image contents such as 
bike, food, etc. have a very low correlation with cyberbullying.}

\section{Cyberbullying detection}

Based on the labeled ground truth data, we next proceed to detect occurrences of cyberbullying in Instagram media sessions. A majority vote criterion was employed to determine whether a media session was cyberbullying/cyberagg- ression or not.   To design and train the classifier, five-fold cross-validation was applied to the data such that 80\% of the data was used for training in each run and 20\% was used for testing, where we report the average.  To achieve a balanced training data set, we over-sampled the minority class of data labeled as cyberbullying (29\% cyberbullying class).

Based on our analysis in Section~\ref{sec:analysis}, we evaluated three types of features, namely those features obtained from the content of comments, those derived from shared media objects, and those obtained from user graph properties of the profile owner, such as the number of followers or followings.  For the text features, we removed characters such as ``!", ``$>$'', etc, as a preprocessing step.  We first focused on unigrams and bigrams.  LIWC categories are derived from unigrams and hence implicitly included as part of this text analysis.  Latent Semantic Analysis (LSA) was employed to reduce the dimensionality of the text analysis, prevent over-fitting and extract the semantics as the features. LSA based on Singular Value Decomposition (SVD) was applied over the unigrams and bigrams. Next, removing stop words such as ``and",``or",``for", etc. was employed, which proved beneficial.  Finally, each feature vector was normalized such that its components sum to one. 

Table~\ref{detection_results} illustrates the performance results for different features and classifiers.  \emph{As we see, linear SVM classifiers based on text-based N-grams with normalization and stop words removal achieve the highest recall (79\%), with precision 71\%.}  Other classifiers were also employed, such as logistic regression, decision tree and AdaBoost, but they did not achieve better results.  

In addition, a variety of non-text features were evaluated, including those features extracted from user behavior (media objects, following, followers), media properties (likes, post time, caption) and image content.  For example, by feeding user graph properties, media statistics and image content to a Na\"{i}ve Bayes classifier, a precision of 48\% and recall of 60\% was achieved. Adding non-text features to the text features for SVM or logistic regression classifiers provided approximately the same results.
\emph{Our conclusion is that the non-text features that we used were not able to meaningfully improve the performance of the cyberbullying detection classifier compared to the text-based features.}  This conclusion is consistent with our earlier analysis from Section \ref{sec:ling}, in which we observed the p-values were small for text-based features, which shows that there was significant differentiation between cyberbullying and non-cyberbullying cases whereas the non-text features in cybebullying and non-cyberbullying samples were not different significantly.


\begin{table*}[htbp!]
\centering
\caption{Cyberbullying detection's classifier performance}
\vspace{1mm}
\begin{tabular}{ l | c | c | c | c  }
\hline 
    \hline
      Features & Classifier & F1-measure & Precision & Recall     \\ \hline
    Unigram-stopword removal  &  linear SVM  &  0.74 &  0.70 & 0.79 \\ \hline 
    Unigram -Bigram- stop word removal &    linear SVM  &  0.75 &  0.77 & 0.73 \\ \hline 
    Unigram -Bigram-stop word removal-Normalization&   linear SVM  &  0.75 &  0.71 & 0.79 \\ \hline 
    Unigram -Bigram-stop word removal-Normalization&   logistic regression  &  0.74 &  0.72 & 0.75 \\ \hline 
 
    \hline
\end{tabular}
\label{detection_results}
\end{table*}

We also applied the same set of features for detecting cyberaggression in Instagram media sessions (38\% cyberaggression). The results are very similar to the cyberbullying 
classifier. Table~\ref{detection_resultsA} shows that both linear SVM and logistic regression provide the same F-1 measure, with recall of 80\% and precision
of 74\% with a logistic regression classifier.  

\begin{table*}[htbp!]
\centering
\caption{Cyberaggression detection's classifier performance}
\vspace{1mm}
\begin{tabular}{ l | c | c | c | c   }
\hline 
    \hline
      Features & Classifier & F1-measure & Precision & Recall     \\ \hline

    Unigram -Bigram-stop word removal-Normalization&  linear SVM  &  0.77 &  0.75 & 0.79 \\ \hline 
 Unigram -Bigram-stop word removal-Normalization&   logistic regression  &  0.77 &  0.74 & 0.80 \\ \hline 
    \hline
\end{tabular}

\label{detection_resultsA}
\end{table*}

\section{Cyberbullying Prediction}
Our next goal is to predict whether cyberbullying will occur at the time an image is posted on Instagram, without benefit of using any subsequent comments in the media session, as we had in the cyberbullying detection.  That means our prediction is based only on the initial posting of the media object, any image features derived from the object, and any properties extant at the time of posting, such as graph features of the profile owner.


To perform the prediction accurately, we need to augment our previous set of labeled media sessions, which were pre-filtered based on text analysis to contain only those with profanity, to also include media sessions lacking in profanity.  This will create a labeled data set that is independent of the content of text comments, namely profanity, and hence usable for prediction, which will not have access to comment contents.
Therefore, we randomly selected 1164 media sessions from the ones with no profanity usage with the criteria of having more than 15 comments and labeled the new set by the same methodology as before. In total 80 contributors worked on the quiz mode, 68 passed, 11 failed and 1 gave up. Labeled data was obtained from 59 trusted labelers as 9 labelers were further filtered out during the labeling process, because they either failed the work mode or rushed through their labeling process. We also labeled the image contents of the new set of media sessions. Table \ref{non_neg_labeling} shows the statistics related to this labeling process. Furthermore, suing the same methodology that we used of cyberbullying detection case, we only consider the media sessions whose labeling had a confidence level more than 60\%.
Overall, this gave us 1,142 labeled media sessions.  

\begin{table}[htbp!]
\centering
\caption{Labeling process statistics. Trusted judgments are the ones made by trusted contributors.}
\vspace{1.5mm}
\begin{tabular}{ p{5cm}  c  }

    \hline
    \hline
   Trusted Judgments  & 5,638  \\  \hline
   Untrusted Judgments & 72 \\ \hline
   Average Test Question Accuracy of Trusted Contributors &	82\% \\ \hline
   Labeled Media Sessions per Hour & 8 \\ \hline
\hline
\end{tabular}

\label{non_neg_labeling}
\end{table}

Table \ref{stats_non_neg} shows the user social graph measurements for the labeled media sessions. We observe that p-values are all less than 0.1 for followers and following, which was not the case for the same features in Table~\ref{graph_stats1}.  This suggests that these features will be more helpful in prediction of cyberbullying than in detection.


\begin{table}[htbp!]
\centering
\caption{Mean values of social graph properties for cyberbullying versus non-cyberbullying samples and cyberaggression versus non cyberaggression, ($^{**}p<0.05,*p<0.1$). }
\vspace{1.8mm}
\begin{tabular}{ l p{1.2cm} p{1.3cm} p{1.2cm}  }
\hline 
    \hline
                         Label  &  *Media objects               & *Following & Followers \\ \hline
   Non-cyberbullying   & 1,157.8      &  721.7 &  **398,283.7  \\ \hline
    Cyberbullying         &    1,198.3            &626.5  &  **465,376.1  \\ \hline
 Non-cyberaggression  &  1,152.6              &  724.4 &   *393,901.6 \\ \hline
    Cyberaggression     &   1,204.3              & 640.3  & *440,403.6    \\ \hline
    \hline
\end{tabular}

\label{stats_non_neg}
\end{table}

The examined features include image contents, media session properties, such as post time and caption, and profile owner social graph properties, such as the number of followers and following and total number of shared media objects.  A similar methodology as the detection was applied for training and testing of the predictor for a total 3,096 media sessions (18\% cyberbullying class).  Table \ref{predresultsb} shows that \emph{cyberbullying incidents can be predicted with recall 76\% and precision 62\%, using only the image contents, media and user meta data based on a Maximum-Entropy (MaxEnt) classifier.  For this augmented complete data set, we observe that  these non-text features provide the main utility for prediction, whereas in detection, the main utility is derived from text-based features over the profanity-based filtered data set.}

In addition, we were interested to explore if prediction could be improved using only a limited set of early comments, not the complete set of comments for a media session.  Hence we also give in  Table \ref{predresultsb} the results of prediction using the first 5, 10 and 15 
comments. We see that as more comments arrive, our predictor is able to improve its precision and recall to 72\% and 77\% respectively using a MaxEnt classifier.  This suggests that text-based features can improve classifier performance.  We hope to leverage this observation in future work to improve predictor performance by utilizing the history of comments from other previously shared media sessions in the prediction.


\begin{table*}[htbp!]
\centering
\caption{Cyberbullying prediction's classifier performance. User properties are followers, following and total number of shared media objects.}
\vspace{1mm}
\begin{tabular}{ l | c | c | c | c  }
\hline 
    \hline
      Features & Classifier & F1-measure & Precision & Recall     \\ \hline
       Image content &  Na\"{i}ve Bayes  &  0.49 & 0.33 & 0.95 \\ \hline 
       User properties, Image content &  Na\"{i}ve Bayes  &  0.56 & 0.42 & 0.84 \\ \hline 
      Image content &  MaxEnt  &  0.33 &  0.21 & 0.85 \\ \hline 
    User properties, Image content &  MaxEnt  &  0.60 &  0.50 & 0.76 \\ \hline 
    User properties, Image content, Post time &  MaxEnt  &  0.63 &  0.56 & 0.74 \\ \hline 
    User properties, Image content, Post time, Caption &  MaxEnt  &  \textbf{0.68} &  \textbf{0.62} & \textbf{0.76} \\ \hline 
   User properties, Image content, Post time, Caption, 5-Comments &  MaxEnt  &  0.71 &  0.69 & 0.72 \\ \hline  
User properties, Image content, Post time, Caption, 10-Comments  &  MaxEnt  & 0.72 &  0.71 & 0.74 \\ \hline  
User properties, Image content, Post time, Caption, 15-Comments  & MaxEnt  & \textbf{0.74}  &  \textbf{0.72}  & \textbf{0.77}  \\ \hline  

    \hline
\end{tabular}
\label{predresultsb}
\end{table*}

Table \ref{predresultsa} provides the results for prediction of cyberaggression media sessions. The F1-measure is 6-9\% lower than for cyberbullying prediction using a MaxEnt classifier, due to significantly lower recall.  

\section{Future Work}

While this paper has advanced the understanding of cyberbullying in a media-based mobile social network, there remain a number of areas for improvement.  One theme for future work is to improve the performance of our classifier.  New algorithms should be considered, such as deep learning and neural networks.  More input features should be evaluated, such as new image features, mobile sensor data, etc.  A limitation of our current classifier is that it is designed only for media sessions that have at least one profanity word.  A more general classifier that can apply to all media sessions is needed.  
Incorporating image features needs to be automated by applying image recognition algorithms.  Temporal behavior of comments for a posted media object shows different behavior for cyberbullying class, which should be taken into account in designing the detection classifier.

 \begin{table*}[htbp!]
\centering
\caption{Cyberaggression prediction's classifier performance. User properties are followers, following and total number of shared media objects.}
\vspace{1mm}
\begin{tabular}{ l | c | c | c | c  }
\hline 
    \hline
      Features & Classifier & F1-measure & Precision & Recall     \\ \hline
         Image content &  MaxEnt  &  0.37 &  0.24 & 0.86 \\ \hline
          User properties, Image content &  MaxEnt  &  0.56  &  0.49  & 0.65  \\ \hline  
      User properties, Image content, Post time &  MaxEnt  &  0.58 &  0.54 & 0.63 \\ \hline 
User properties, Image content, Post time, Caption &  MaxEnt  &  \textbf{0.62} &  \textbf{0.61} & \textbf{0.63} \\ \hline
  User properties, Image content, Post time, Caption, 5-Comments &  MaxEnt  &  0.63  &  0.68  & 0.59  \\ \hline  
User properties, Image content, Post time, Caption, 10-Comments  &  MaxEnt  & 0.64  &  0.69  & 0.60  \\ \hline  
User properties, Image content, Post time, Caption, 15-Comments  & MaxEnt  &  \textbf{0.65}  &  \textbf{0.7}   &  \textbf{0.61} \\ \hline 
    \hline
\end{tabular}
\label{predresultsa}
\end{table*}

In this work we have only considered the image content and image and user metadata for prediction of cyberbullying.  However, based on the improvement seen in using a small number of text comments, we think that considering the commenting history of users in previously shared media can prove to be useful. 

Another theme for future work is to obtain greater detail from the labeling surveys.  Our experience was that streamlining the survey improved the response rate, quality and speed.  However, we desire more detailed labeling, such as for different roles in cyberbullying -- identifying and differentiating the role of a victim's defender, who may also spew negativity, from a victim's bully or bullies.

\section{Conclusions}

This paper makes the following major contributions: (1) an appropriate definition of cyberbullying that incorporates both frequency of negativity and imbalance of power applied in large-scale labeling, and is differentiated from cyberaggression; (2) cyberbullying is studied in the context of a media-based social network, incorporating both images and comments in the labeling; (3) a detailed analysis of the distribution results of labeling of cyberbullying incidents is presented, including a correlation analysis of cyberbullying with other factors derived from images, text comments, and social network meta data; (4) multi-modal classification results are presented for cyberbullying detection as well as
prediction that incorporate a variety of features to identify cyberbullying incidents.

The major findings of this paper comprise the following results.  First, we found that labelers are mostly in agreement about what constitutes cyberbullying and cyberaggression in Instagram media sessions.  Second, we found a significant number of media sessions containing profanity and cyberaggression that were not labeled as cyberbullying, suggesting that classifiers for cyberbullying must be more sophisticated than mere profanity detection.  Third, media sessions with very high percentage of negativity above 60-70\% actually correspond to a lower likelihood of cyberbullying.  Fourth, media sessions with cyberbullying exhibit more frequent commenting.  Fifth, users of media sessions containing cyberbullying demonstrate a lower number of likes per post.  Sixth, cyberbullying has a higher probability of occurring when media sessions contain certain linguistic categories such as death, appearance, religion and sexuality content.  Seventh, certain image contents such as ``drug" are highly related to cyberbullying while other image categories such as ``tattoo" or ``food" are not.  Eighth, a linear SVM classifier can detect cyberbullying with 79\% recall and  71\% precision using only text features, where non-text features were not helpful.  Ninth, however, non-text features such as image and user meta data were central to cyberbullying prediction, where a MaxEnt classifier achieved 76\% recall and 62\% precision.

\bibliographystyle{abbrv}
\bibliography{homa}

\end{document}